\documentclass[12pt,preprint]{aastex}           

\usepackage{psfig} 
\bibpunct[; ]{(}{)}{;}{a}{}{,}

\begin{document}

\title{The Anatomy of Star Formation in NGC~300}
 
\author{G. Helou,\altaffilmark{1}
H. Roussel,\altaffilmark{2}\\
P. Appleton,\altaffilmark{1}
D. Frayer,\altaffilmark{1}
S. Stolovy,\altaffilmark{1}
L. Storrie-Lombardi,\altaffilmark{1}
R. Hurt,\altaffilmark{1}
P. Lowrance,\altaffilmark{1}\\
D. Makovoz,\altaffilmark{1}
F. Masci,\altaffilmark{1}
J. Surace,\altaffilmark{1}
K.D. Gordon,\altaffilmark{3}
A. Alonso-Herrero,\altaffilmark{3}
C.W. Engelbracht,\altaffilmark{3}\\
K. Misselt,\altaffilmark{3}
G. Rieke,\altaffilmark{3}
M. Rieke,\altaffilmark{3}
S.P. Willner,\altaffilmark{4}
M. Pahre,\altaffilmark{4}
M.L.N. Ashby,\altaffilmark{4}\\
G.G. Fazio,\altaffilmark{4}
H.A. Smith\altaffilmark{4}
}

\altaffiltext{1}{\scriptsize SIRTF Science Center, California Institute of Technology, M.S. 220-6, Pasadena, CA 91125}

\altaffiltext{2}{\scriptsize California Institute of Technology, M.S. 320-47, Pasadena, CA 91125}

\altaffiltext{3}{\scriptsize Steward Observatory, University of Arizona, Tucson, AZ 85721}

\altaffiltext{4}{\scriptsize Harvard-Smithsonian Center for Astrophysics, 60 Garden St., Cambridge, MA 02138}

\begin {abstract}
The {\it Spitzer Space Telescope}  was used to study the mid-infrared 
to far-infrared properties of NGC~300, and to compare dust emission to
H$\alpha$ to elucidate the heating of the ISM and the star formation
cycle at scales $< 100$\,pc. The new data allow us to discern clear
differences  in the spatial distribution of $8\,\mu$m dust emission with
respect to $24\,\mu$m dust and to \ion{H}{2} regions traced by the
H$\alpha$ light. The $8\,\mu$m emission highlights the rims of \ion{H}{2}
regions, and the $24\,\mu$m emission is more strongly peaked in star
forming regions than $8\,\mu$m. We confirm the existence and approximate
amplitude of interstellar dust emission at $4.5\,\mu$m, detected
statistically in Infrared Space Observatory (ISO) data, and conclude
it arises in star forming regions. When averaging over regions larger
than $\sim 1$\,kpc, the ratio of H$\alpha$ to Aromatic Feature emission
in NGC~300 is consistent with the values observed in disks of spiral galaxies.
The mid-to-far-infrared spectral energy distribution of dust
emission is generally consistent with pre-Spitzer models.
\end {abstract}
 

\section {Introduction}

\begin{figure}[!ht]
\vspace*{-2.5cm}
\resizebox{15.5cm}{!}{\plotone{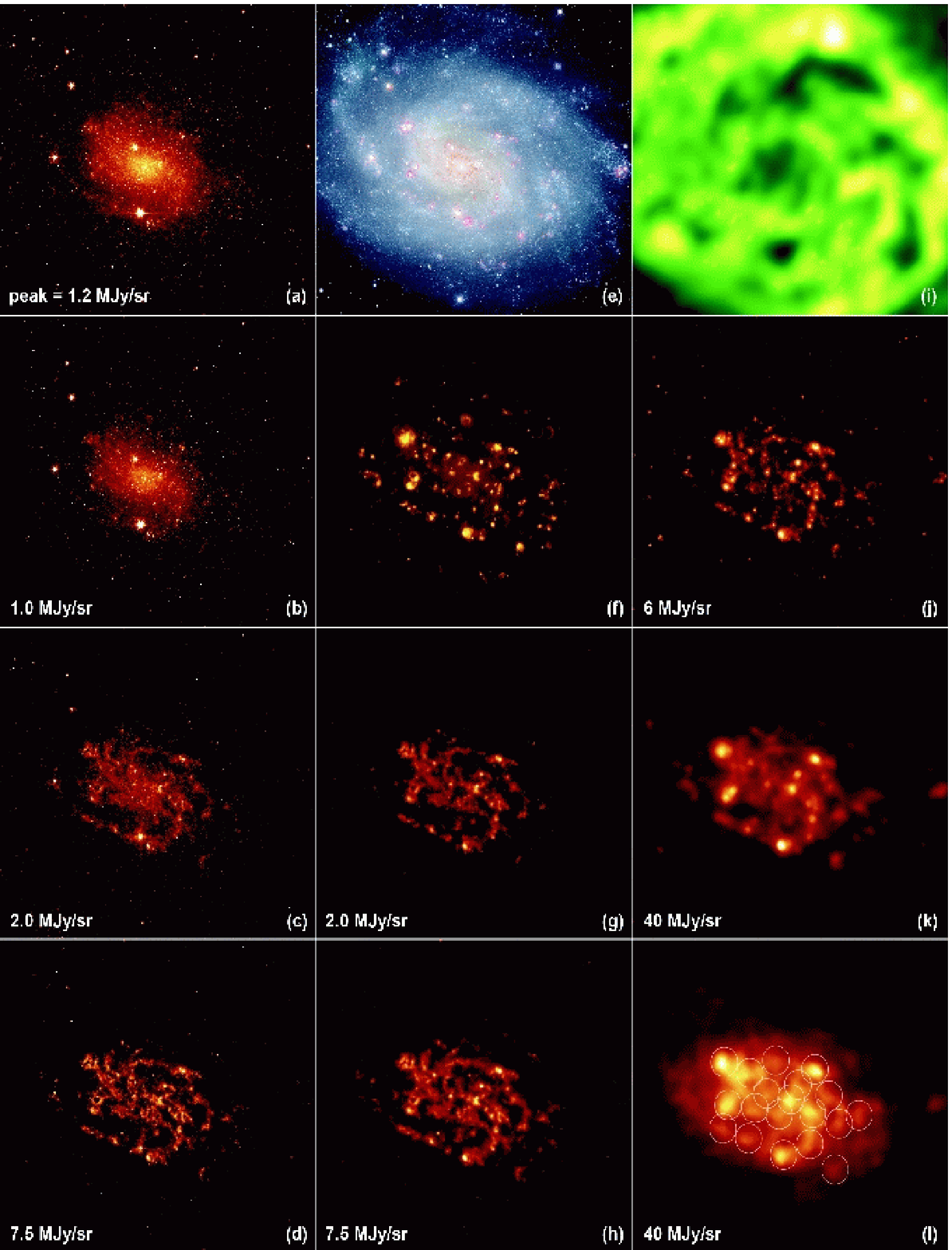}}
\caption[]{Images of NGC~300 obtained with the {\it Spitzer
Space Telescope} and other sources.  The left-most column shows observed maps at
(a) $3.6\,\mu$m,
(b) $4.5\,\mu$m,
(c) $5.8\,\mu$m,
(d) $8\,\mu$m.  The center column shows
(e) BVRH$\alpha$ composite image (credit MPG/ESO),
    where B is coded in blue, V in green, R and H$\alpha$ in red;
(f) H$\alpha$ map described in Section~\ref{data};
(g) dust map at $5.8\,\mu$m;
(h) dust map at $8\,\mu$m.  The right-most column shows
(i) \ion{H}{1} map from \citet{Puche}; and observed maps at
(j) $24\,\mu$m;
(k) $70\,\mu$m;
(l) $160\,\mu$m, with photometric apertures
used in Figure~\ref{fig:sed} superposed. Each frame contains
the peak surface brightness in the background-subtracted image; all images
are in logarithmic scale, except the $160\,\mu$m image which is linear.
North is about $40\degr$ clockwise from vertical up,
and each frame is approximately $20^{\prime}$ on a side.
}
\label{fig:panel1}
\end{figure}

NGC~300 is a SA(s)d galaxy in the Sculptor Group of galaxies, at a distance
of about 2.1~Mpc \citep{Freedman}, viewed at an inclination of about $50\degr$
\citep{Puche}.
Its total luminosities in the blue band (3855--4985\AA) and in the
far-infrared (42.5--$122.5\,\mu$m, estimated from Spitzer data) are
$\sim 3.3 \times 10^8$\,L$_{\sun~bol}$ and $2.2 \times 10^8$\,L$_{\sun~bol}$
respectively. Its $L_{\rm FIR} / L_{\rm B}$ ratio is thus very close
to the average ratio for the blue magnitude-limited sample studied by
\citet{Thuan}.
Due to its large angular extent and low surface brightness,
no reliable total radio continuum flux measurement exists (only discrete
sources are detected). NGC~300 has a striking
appearance in the visible and in H$\alpha$ due to several
\ion{H}{2} regions with nearly circular shapes and various degrees
of filling-in \citep[][see Figure~\ref{fig:panel1}]{Deharveng, Hoopes}.
Its large HI envelope \citep{Puche} extends well beyond the
visible image. \citet{Pannuti} reported a total of 44 SNR candidates,
evidence that the current star formation activity manifested as
\ion{H}{2} regions has been on-going for tens of millions of years
\citep[see also][]{Butler}.
Because of its proximity, NGC~300 allows the Spitzer Space Telescope
\citep{Werner} to compare dust emission to other ISM components and
discern the interplay between the ISM and the star
formation cycle at scales $< 100$\,pc.

\section{Observations and Data Reduction}
\label{data}

The IRAC \citep{Fazio} observations of NGC~300 are 12-second frames
and map the galaxy in approximately half array spacings, yielding a total
time per sky position of 48\,s. They were reduced with the standard Spitzer
Science Center data reduction pipeline (version 9.5).
Due to the readout
of bright point sources, the signal was reduced (pulled-down) in some
columns of the array\footnote{see the Spitzer Observer's Manual at
{\it http://ssc.spitzer.caltech.edu/documents/som/}.};
a correction measured from the map was applied.
Persistent images left by bright sources were found by median-combining
all the dithered positions together and identifying any remaining sources.
These objects were removed from individual frames. The data were then
combined into a mosaic using a cosmic ray rejection and a background matching
applied between overlapping fields of view. The relative photometric
uncertainty is of the order of 5\%, and the absolute uncertainty is 10\%;
uncertainty tied to the angular sizes of source and measuring aperture
contributes up to 15\%.

Images of NGC~300 at 24, 70, and $160\,\mu$m were obtained with the MIPS
Instrument \citep{Rieke} in the scan-map mode.
The final mosaics have a total exposure time of approximately 160, 80, and
16\,s per point at 24, 70, and $160\,\mu$m.  The MIPS images
were reduced using the MIPS Instrument Team Data Analysis Tool
\citep{Gordon} as described by \citet{Engelbracht}. The
uncertainties on the final absolute calibrations are estimated at
10\%, 20\%, and 20\% for the 24, 70, and $160\,\mu$m data,
respectively. The 70 and $160\,\micron$ images exhibit linear streaks
along the scan direction which are residual instrumental artifacts due
to the time dependent responsivity of the Ge detectors, and affect
the photometry in large apertures.  

The H$\alpha$ (6563\AA) $+$ [\ion{N}{2}] (6583 and 6548\AA)
map was derived from images posted in NED by \citet{Larsen}. The narrowband
image containing H$\alpha$ emission and the R-band image were aligned,
and rescaled to subtract the stellar continuum, deriving the scaling factor
from aperture photometry on 32 bright stars. Residuals from saturated or
improperly subtracted stars were masked out, and the final image was
regridded and rotated to match the IRAC reference frame.

The Spitzer and H$\alpha+$[\ion{N}{2}] images are shown in
Figure~\ref{fig:panel1}. The images contain three very bright Milky Way
stars directly superposed on the disk. We masked them out in IRAC and
$24\,\mu$m images, as well as additional point sources as described in
Section~\ref{dust_colors} below (Figure~\ref{fig:panel2}b).
Foreground stars will be removed more rigorously 
in a later paper. Fainter, indistinguishable stars are
unlikely to affect significantly the results presented here.

\section{Physical Content in Mid-Infrared Images}

\begin{figure}[!ht]
\resizebox{10cm}{!}{\rotatebox{90}{\plotone{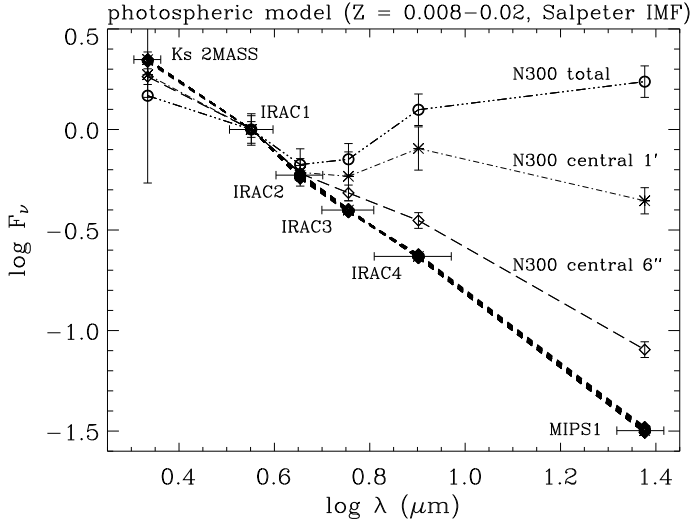}}}
\vspace*{-0.5cm}
\caption[]{SED decomposition: 
The SEDs of NGC~300 in three different apertures are
shown, with all fluxes normalized to 3.6\,$\mu$m: 
the nuclear stellar cluster (clearly contaminated by dust emission
judging from morphology in the IRAC maps), the inner arcmin, where dust
contribution is relatively small, and the whole galaxy. 
The thick dashed lines trace model photospheric SEDs from 2 to 24\,$\mu$m, derived
from Starburst99 for a range of star formation histories and two metallicities,
and transformed to IRAC filter fluxes. 
The horizontal error bars represent the filter widths at half maximum transmission.
The Ks fluxes are lower limits, because the 2MASS map is not sensitive to
the diffuse emission.
}
\label{fig:stel_model}
\end{figure}

Whereas at $\lambda \ge 24\,\mu$m the emission is dominated by interstellar
dust, the mid-infrared marks the transition from stellar to interstellar
emission. The appearance of the galaxy shifts from resembling the stellar
disk in the $3.6\,\mu$m image to an ISM-like appearance at $8\,\mu$m.
We assume conservatively the $3.6\,\mu$m emission is all due to stars,
and extrapolate this component to longer wavelengths using stellar
population modeling from Starburst99 \citep{Leitherer}; the stellar
Spectral Energy Distribution (SED) longward of $2\,\mu$m depends very
little on star formation history or metallicity, as illustrated in
Figure~\ref{fig:stel_model}. The decomposition amounts to scaling the
$3.6\,\mu$m map by
0.596, 0.399, 0.232 and 0.032 respectively for 4.5, 5.8, 8 and $24\,\mu$m,
and subtracting these from the observed maps pixel by pixel
to yield what we will refer to as the ''dust maps" shown in
Figure~\ref{fig:panel1}. Based on this decomposition, the $24\,\mu$m map
is probably $\ge 98$\% interstellar emission globally, and $\ge 93$\%
interstellar in the inner arcmin, where the stellar contribution is the
greatest. The $8\,\mu$m map is $\ge 81$\% interstellar emission.

This method potentially overestimates the dust component by underestimating
the redder SED of very low mass stars. However the model reproduces
adequately the $4.5\,\mu$m/$3.6\,\mu$m color of the nuclear stellar cluster,
of the emission averaged in the inner arcmin (Figure~\ref{fig:stel_model}),
and of the diffuse emission outside several disk \ion{H}{2} regions.

On the other hand, the method probably underestimates the dust component
by assuming that the $3.6\,\mu$m map contains only photospheric emission,
ignoring evidence that the hot dust component stretches down to
$\lambda < 3\,\mu$m \citep{Bernard, Lu, Hunt}. From the ISO data in Figure~6
of Lu et al. (2003), this underestimate may reach a factor of up to 2.6,
10--20\% and 3--4\%,  respectively at 4.5, 5.8 and $8\,\mu$m. In view
of these opposing potential biases, our conservative approach should be
reliable qualitatively for NGC~300, in spite of localized departures due
to variations in stellar population colors or dust extinction.

\subsection{Interstellar Dust Emission}
\label{dust_colors}

\begin{figure}[!ht]
\resizebox{10cm}{!}{\rotatebox{90}{\plotone{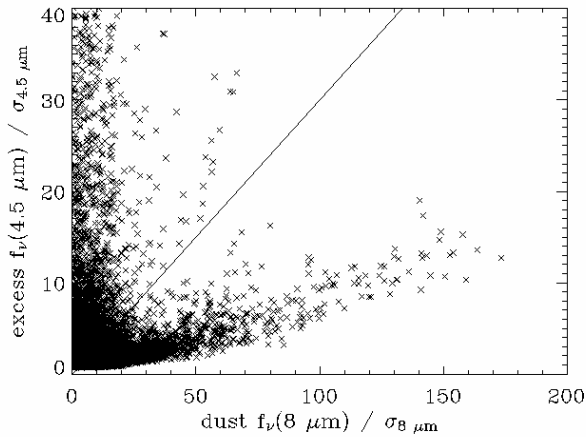}}}
\vspace*{-0.5cm}
\caption[]{
Scatter plot of excess $f_{\nu}(4.5\,\mu{\rm m})$ versus dust-only
$f_{\nu}(8\,\mu{\rm m})$, normalized by the noise in the original
maps. Individual pixels are plotted, after convolution of all maps
to a common angular resolution.
The correlation branch was isolated by selecting pixels below
the solid line (corresponding to Figure~\ref{fig:panel2}a).
The uncorrelated emission, extending to much higher brightnesses,
was cut here for clarity. The scatter plot versus H$\alpha$, not
included, shows a similar correlation branch, most pixels of which
map into the 8\,$\mu$m correlation branch.
}
\label{fig:corr_excess_irac2}
\vspace*{-10cm}
\end{figure}

In order to constrain the origin of the $4.5\mu$m non-stellar emission,
we identified pixels where the excess is above the $3\sigma$ significance
threshold in the $4.5\mu$m ''dust map", and examined their colors and
locations. In scatter plots of these pixel values {\it vs.} H$\alpha$
or {\it vs.} $8\,\mu$m dust emission, a correlation branch is evident
in both cases with moderate slopes in
$f_{\nu} (4.5\,\mu m)/f_{\nu} (8\,\mu m)$ and $f_{\nu} (4.5\,\mu m)$/H$\alpha$
(Figure~\ref{fig:corr_excess_irac2}),
as well as a branch with very steep, almost vertical slopes. The pixels
in the latter branch are strongly associated with the $3.6\,\mu$m emission,
including the brightest foreground star and a large number of point sources
spread almost uniformly accross the whole map. We use this excess map as
a mask to efficiently remove foreground stars, with additional hand
masking of residues around very bright stars.

On the other hand, the branches with moderate $4.5\,\mu$m/$8\,\mu$m and
$4.5\,\mu$m/H$\alpha$ ratios map into bright star formation regions, as
might be expected (Figure~\ref{fig:panel2}a). The global ratio
$\nu f_{\nu} (4.5\,\mu$m)/$\nu f_{\nu} (8\,\mu$m) for the selected dust
emission amounts to  $\sim 4$\%. However, summing the pixels in star
forming regions with large $4.5\,\mu$m excess, dust alone accounts for
17\% of the total $4.5\,\mu$m emission, probably more if dust contribution
to the $3.6\,\mu$m emission were accounted for. Two uncertainties remain
which will be addressed in future papers: Whether redder SEDs of young
stars affect significantly the results in star forming regions; and
whether the $4.5\mu$m dust associates more closely with ionized regions
or with Photo-Dissociation Regions (PDR).

The dust maps at 5.8 and $8\,\mu$m should be dominated by the Aromatic Features
in Emission (AFE) proposed by \citet{Puget}) to explain IRAS data, and well
studied with ISO \citep{Boulanger98, Helou00, Lu}. The value and variability
of the ratios between these bands are of considerable interest in constraining
the physical state of the emitters \citep[e.g.][]{Frees}. \citet{Lu} found
the AFE profiles among galaxies constant to within measurement uncertainty
of 15 to 25\% in the range 5 to 12$\mu$m; the Spitzer images can extend this
finding to much greater accuracy, and could point to variations within disks
on a variety of scales.

We find that  $\nu f_{\nu} (5.8\,\mu$m)/$\nu f_{\nu} (8\,\mu$m) for dust
emission alone is constant to within the uncertainties, with a value of
$0.50 \pm 0.09$ ($3 \sigma$) over most of the area where surface brightness
is above the $6 \sigma$ level in each of the $5.8\,\mu$m and $8\,\mu$m dust maps.

\subsection{Comparison to H$\alpha$}

\begin{figure}[!ht]
\vspace*{-1cm}
\resizebox{15cm}{!}{\plotone{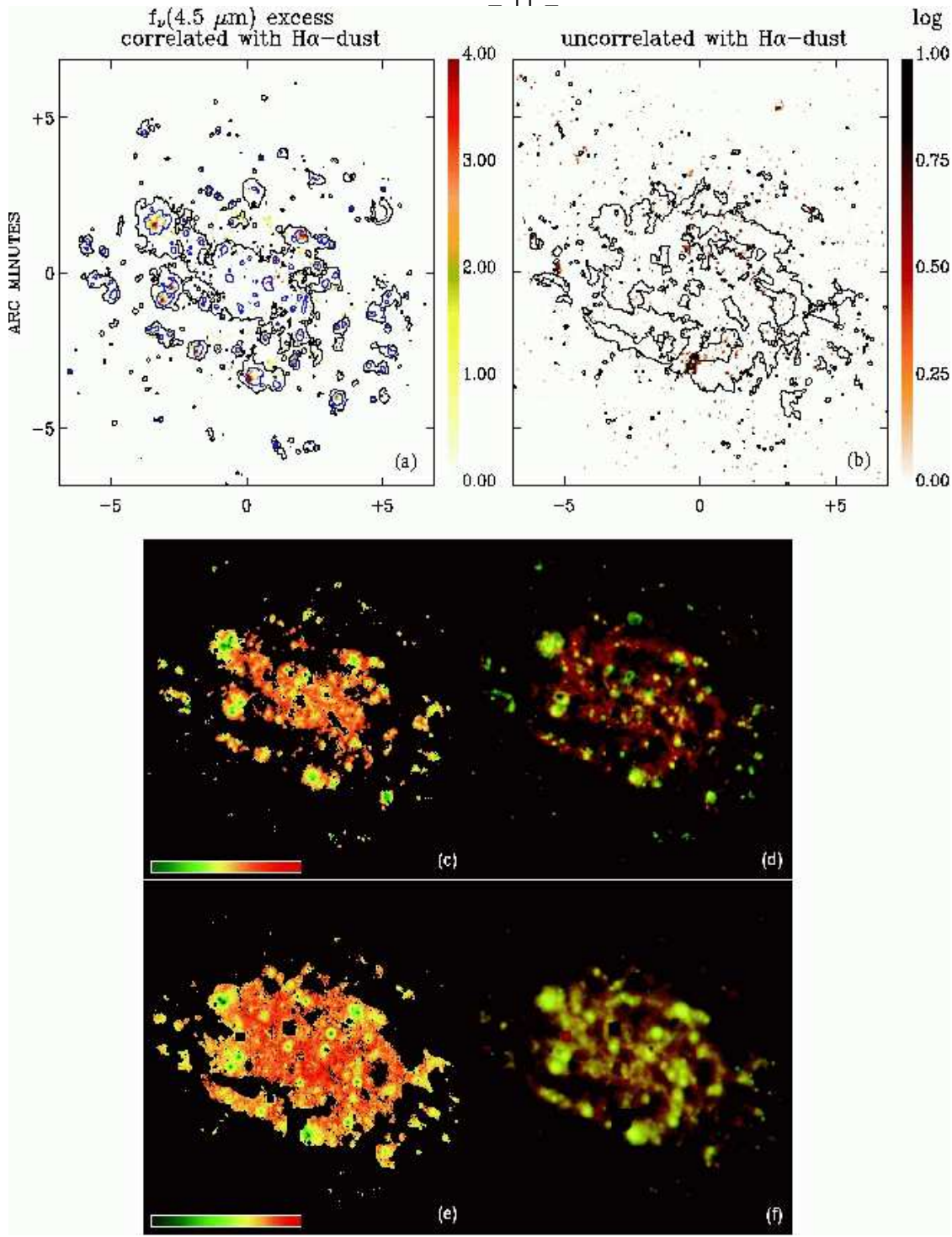}}
\caption[]{
{\bf Top:}~Distribution of excess 4.5\,$\mu$m emission,
$f_{\nu}^e$(4.5\,$\mu$m), in units of the noise in the initial
4.5\,$\mu$m map, $\sigma_{4.5}$.
{\bf (a):} Part of $f_{\nu}^e$(4.5\,$\mu$m) which is correlated with
8\,$\mu$m and H$\alpha$ (950 pixels $\in [3;19]~ \sigma_{4.5}$.
{\bf (b):} Part of $f_{\nu}^e$(4.5\,$\mu$m) which is not correlated
with 8\,$\mu$m nor H$\alpha$, in logarithmic scale
(4583 pixels $\in [3;8150]~ \sigma_{4.5}$).
The contours are H$\alpha$ isophotes (a) and the $4 \sigma$
isophote of the 8\,$\mu$m dust map (b).
{\bf Bottom:} Comparison of H$\alpha$ and dust emission: 
(c) log($f_{\nu}(8\,\mu$m) / H$\alpha$) (95\% of pixels span 1.2\,dex),
showing low ratios inside prominent \ion{H}{2} regions;
(e) log($f_{\nu}(8\,\mu$m) / $f_{\nu}(24\,\mu$m))
(95\% of pixels span the linear range [0.6; 2.2]),
showing again the distinctly low color ratios of \ion{H}{2} regions;
(d) $8\,\mu$m (red) and H$\alpha$\, (green) composite image;
(f) $8\,\mu$m (red) and $24\,\mu$m (green) composite image.
}
\label{fig:panel2}
\vspace*{-5cm}
\end{figure}

Figure~\ref{fig:panel2}c shows a comparison between H$\alpha$ and $8\,\mu$m
maps at a resolution of
$2 \times {\rm FWHM} \sim 5.15^{\prime\prime} \sim 50$\,pc.
In the outer disk, several \ion{H}{2} regions as well as more complex
structures are visible in the H$\alpha$ map, with the corresponding
$8\,\mu$m profile flatter and sometimes significantly offset from the
H$\alpha$ peak. In spite of the complex structure, Figure~\ref{fig:panel2}c-d
clearly shows the $8\,\mu$m emission highlighting the rims
of \ion{H}{2} regions. The ratio $8\,\mu$m/H$\alpha$ goes up by a factor
five to ten from inside to just outside the \ion{H}{2} region. This ratio
may increase by another factor of 2--4 in diffuse regions, but this is
uncertain because of limited sensitivity and inaccurate continuum subtraction
in the H$\alpha$ map. While this behavior is not a complete surprise,
the Spitzer data are the first to show it so clearly in a nearby galaxy.
\citet{Boulanger90}, \citet{Sellgren},
\citet{Giard} and others have reported sharp boundaries to the AFE, 
with peaks on molecular cloud surfaces outside ionized regions, and related 
this to a combination of UV excitation and photo-processing of the AFE carriers.
\citet{Helou01} also speculated that AFE arise primarily in PDRs based on
physical arguments rather than direct observations. Inside the \ion{H}{2}
regions, the Aromatics are depressed probably because of destruction by
the ionizing UV \citep{Boulanger90}.

The sharp contrasts near \ion{H}{2} regions between $8\,\mu$m and H$\alpha$
are only a small-scale property. Summing over scales larger than about 1\,kpc,
the ratio between the two tracers is constant to about 40\%.
The global ratio of $8\,\mu$m to unextincted H$\alpha$ is roughly
30\% smaller than the average measured in disks of spiral galaxies by
\citet{Roussel01a}, which had itself a dispersion of 50\% among disks.
For this comparison, the total H$\alpha+$[\ion{N}{2}] flux from \citet{Hoopes}
was corrected for an extinction $A$(H$\alpha) = 0.37$ and an
H$\alpha$/(H$\alpha+$[\ion{N}{2}]) ratio of 0.91, averaged from
values of individual \ion{H}{2} regions in \citet{Webster}.
The $8\,\mu$m IRAC flux was converted to $7\,\mu$m flux in the ISO filter
LW2 assuming the same spectral shape as in the disk of M\,83 \citep{Roussel01b},
which was verified to reproduce accurately the $5.8\,\mu$m/$8\,\mu$m
dust ratio of NGC~300.

The star formation rate in the disk of NGC~300 is
0.08--0.11~M$_{\sun}$\,yr$^{-1}$ from the $8\,\mu$m emission
\citep[as calibrated for solar-metallicity spiral disks by][]{Roussel01a},
and $\sim 0.14$~M$_{\sun}$\,yr$^{-1}$ from H$\alpha$
\citep[as calibrated by][]{Kennicutt}.
The agreement points to $8\,\mu$m under these conditions as a viable proxy
for ionizing flux, and therefore for on-going massive star formation rate.
Since the association between $8\,\mu$m and H$\alpha$ is not at the
atomic process level however, it will be subject to geometric and other
effects, and the use of $8\,\mu$m as such a proxy is unlikely to be valid
under all conditions.

Figure~\ref{fig:panel2}e-f also shows the $24\,\mu$m emission
is more peaked than $8\,\mu$m in star forming regions. The spatial resolution
is insufficient to distinguish whether PDRs or ionized regions are more closely
associated with the $24\,\mu$m peaks. This confirms earlier suggestions
based on global SED analysis that $24\,\mu$m traces heating by the youngest
and most massive stars in a galaxy \citep{Helou00bis}.

\section{Mid-IR to Far-IR SED}

\begin{figure}[!ht]
\resizebox{12cm}{!}{\plotone{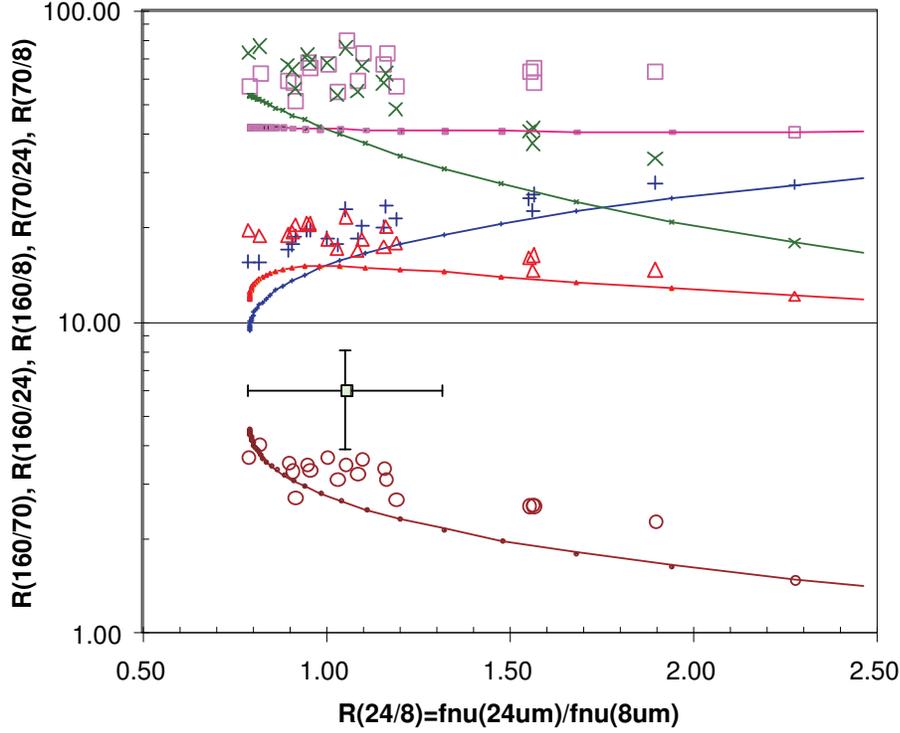}}
\caption[]{Color-color plot for all Spitzer bands at $\lambda \ge 8\,\mu$m,
for the regions measured with an aperture
corresponding to $\sim 1$\,kpc in the plane of NGC~300,
with the \citet{Dale02} model predictions superposed as curves.
Each curve has one symbol at x=2.28 to indicate the corresponding color ratio.
We corrected the
$8\,\mu$m fluxes by a factor of $\sqrt{0.69}$
since the aperture is intermediate in size between the nominal point-source
aperture and infinitely extended scales.  The one point with error bars
illustrates the uncertainties of 25\%
on the x-axis and 35\% on the y-axis.
Squares are for $160\,\mu$m/$8\,\mu$m; X signs are for $160\,\mu$m/$24\,\mu$m;
plus signs are for $70\,\mu$m/$8\,\mu$m; triangles are for $70\,\mu$m/$24\,\mu$m;
and circles are for $160\,\mu$m/$70\,\mu$m.
The systematic offset at the longer wavelength bands is discussed in the text.} 
\label{fig:sed}
\end{figure}

In order to study the SED at $\lambda \ge 8\,\mu$m, we identified nineteen
emission regions in the dust maps, most centered on a local peak
(Figure~\ref{fig:sed}l).
After smoothing all maps to the resolution of $160\,\mu$m, we measured the
emission of each region in a circular aperture
$2.5 \times {\rm FWHM}(160\,\mu{\rm m}) = 95^{\prime\prime} \sim 1$\,kpc
in diameter. The resulting photometry is plotted in Figure~\ref{fig:sed} as
various color ratios against $f_{\nu}(8\,\mu{\rm m})/f_{\nu}(24\,\mu{\rm m})$,
and compared to the ratios expected from pre-Spitzer model SEDs of galaxies
\citep{Dale02}. These are single-parameter models of dust emission integrated
over whole galaxies, and incorporate a power-law distribution of dust over
heating intensities. The data and model agree in the {\it trends}, though
there are systematic offsets in the values. These offsets are most simply
understood as a systematic discrepancy of $\sim 30$\% at $160\,\mu$m and
$\sim 15$\% at $70\,\mu$m between model and measurement, assuming reliable
photometry at 8 and $24\,\mu$m. In view of the simplicity of the models,
photometric uncertainties (especially non-linearity effects)
and source extent, the agreement between model
and data is quite satisfactory, and suggests that beyond a certain scale
portions of galaxies may approach full galaxies in their behavior.

The $160\,\mu$m and less so the $70\,\mu$m maps display a striking large halo
which echoes several features of the HI envelope. It is unclear at this point
whether this reflects unusually cold material in the outer disk, or whether the
larger $160\,\mu$m beam is simply more sensitive to low surface brightness
features.

\section{Summary}

The Spitzer Space Telescope data allow us more clearly than ever to dissect the
interstellar dust emission components in NGC~300 and to relate them to 
star formation.  The $8\,\mu$m AFE highlights the rims of the \ion{H}{2} regions
defined by H$\alpha$, confirming it is more closely associated
with PDRs than with ionized regions.
When averaging over regions larger than $\sim 1$\,kpc, the ratio of
H$\alpha$ to AFE emission in NGC~300 is consistent with the disk value
observed in other galaxies. This confirms AFE as a
convenient massive star formation rate estimator in disks of galaxies.
The mid-IR spectral signature of the AFE is invariant as measured by the ratio
$f_{\nu}(5.8\,\mu{\rm m})/f_{\nu}(8\,\mu{\rm m})$, which is  
essentially constant at $\sim 50$\,pc resolution at the $\sim 20$\%
($3 \sigma$) measurement accuracy over most of the disk of NGC~300.
The $24\,\mu$m emission is more strongly
peaked than $8\,\mu$m in star forming regions. Interstellar dust emission
at $4.5\,\mu$m is confirmed at about the expected amplitude; for the
first time, its spatial distribution associates it clearly with star
forming regions.  The data cannot distinguish
between PDRs and \ion{H}{2} regions as the primary origin for either 4.5 or
$24\,\mu$m dust emission.
The SED models of \citep{Dale02} apply to $\sim$kpc-sized portions in the disk
to within the current photometry and calibration uncertainties of 20--30\%.

\acknowledgements
The Spitzer Space Telescope is operated by the Jet Propulsion Laboratory,
California Institute of Technology, under contract with the National
Aeronautics and Space Administration.
This research has made use of the NASA/IPAC Extragalactic Database which is
operated by JPL/Caltech, under contract with NASA.

\clearpage

\renewcommand{\baselinestretch}[0]{1.}

\clearpage

\vspace*{-1cm}

\clearpage

\clearpage

\end{document}